\documentclass[fleqn,10pt]{wlscirep}
\title{Pure spin-Hall magnetoresistance in Rh/Y$_3$Fe$_5$O$_{12}$ hybrid}
\author[1]{T. Shang}
\author[1,*]{Q. F. Zhan}
\author[2]{L. Ma}
\author[1]{H. L. Yang}
\author[1]{Z. H. Zuo}
\author[1]{Y. L. Xie}
\author[1]{H. H. Li}
\author[1]{L. P. Liu}
\author[1]{B. M. Wang}
\author[3]{Y. H. Wu}
\author[4,$\dag$]{S. Zhang}
\author[1,$\ddag$]{Run-Wei Li}
\affil[1]{Key Laboratory of Magnetic Materials and Devices \& Zhejiang Province Key Laboratory of Magnetic Materials and Application
Technology, Ningbo Institute of Material Technology and Engineering, Chinese Academy of Sciences, Ningbo, Zhejiang 315201, P. R. China}
\affil[2]{Department of Physics, Tongji University, Shanghai, 200092, P. R. China}
\affil[3]{Department of Electrical and Computer Engineering, National University of Singapore, 4 Engineering Drive 3 117583, Singapore}
\affil[4]{Department of Physics, University of Arizona, Tucson, Arizona 85721, USA}
\affil[*]{zhanqf@nimte.ac.cn}
\affil[$\dag$]{zhangshu@email.arizona.edu}
\affil[$\ddag$]{runweili@nimte.ac.cn}

\begin{abstract}
We report an investigation of anisotropic magnetoresistance (AMR) and anomalous Hall resistance (AHR) of Rh and Pt thin films sputtered on epitaxial Y$_3$Fe$_5$O$_{12}$ (YIG) ferromagnetic insulator films. For the Pt/YIG hybrid, large spin-Hall magnetoresistance (SMR) along with a sizable conventional anisotropic magnetoresistance (CAMR) and a nontrivial temperature dependence of AHR were observed in the temperature range of 5-300 K. In contrast, a reduced SMR with negligible CAMR and AHR was found in Rh/YIG hybrid. Since CAMR and AHR are characteristics for all ferromagnetic metals, our results suggest that the Pt is likely magnetized by YIG due to the magnetic proximity effect (MPE) while Rh remains free of MPE. Thus the Rh/YIG hybrid could be an ideal model system to explore physics and devices associated with pure spin current.
\end{abstract}
\begin{document}

\maketitle

\section*{Introduction}

The studies of magnetic insulator-based spintronics have recently generated great interest due to their segregated characteristic of spin current from charge current~\cite{Wu2013}. The interplay between spin and charge transports in nonmagnetic metal/ferromagnetic insulator (NM/FMI) hybrids gives rise to various interesting phenomena, such as spin injection~\cite{Ohno1999,Jedema2001}, spin pumping~\cite{Heinrich2011,Rezende2012,Kajiwara2010}, and spin Seebeck~\cite{Uchida2008,Uchida2010}. The previous investigations on NM/FMI hybrids, e.g., Pt/Y$_3$Fe$_5$O$_{12}$ (Pt/YIG), also demonstrated a new-type of magnetoresistance~\cite{Miao2014,Althammer2013,Isasa2014,Lin2014,Hahn2013} in which the resistivity of films, $\rho$, has an unconventional angular dependence, namely,
\begin{equation}
\rho = \rho_0 - \Delta \rho \left[\hat{\bf m} \cdot (\hat{\bf z}\times \hat{\bf j}) \right]^2
\end{equation}
where $\hat{\bf m}$ and $\hat{\bf j}$ are unit vectors in the directions of the magnetization and the electric current, respectively, and $\hat{\bf z}$ represents the normal vector perpendicular to the plane of the film; $\rho_0 $ is the zero-field resistivity. The above angular-dependent resistivity has been named as the spin-Hall magnetoresistance (SMR) in order to differentiate from the conventional anisotropic magnetoresistance (CAMR) in which $\rho = \rho_0 + \Delta \rho (\hat{\bf m} \cdot \hat{\bf j})^2$. A theoretical model outlined below has been proposed to explain the SMR. An electric current (${\bf j}_e$) induces a spin current due to the spin-Hall effect and in turn, the induced spin current, via inverse spin-Hall effect, generates an electric current whose direction is opposite to the original current\cite{Nakayama2013, Chen2013, Hirsch1999, Wunderlich2005, Kato2004, Tatara2006, Kajiwara2006, Kimura2007}. Thus , the combined spin-Hall and inverse spin-Hall effects lead to an additional resistance in bulk materials with spin-orbit coupling (SOC). However, in an ultra-thin film, the spin current could be either reflected back at the interface or absorbed at the interface through spin transfer torque. In the former case, the total spin current in the metal layer is reduced and thus the additional resistance is minimized. The spin current reflection is strongest when the magnetization direction $\hat{\bf m}$ of the ferromagnetic insulator is parallel to the spin polarization $\hat{\bf z}\times \hat{\bf j}$ of the spin current, leading to the resistive minimum as described in Eq.~(1)~\cite{Nakayama2013,Chen2013}.

However, the magnetic proximity effect (MPE), in which a non-magnetic metal develops a sizable magnetic moment in the close vicinity of a ferromagnetic layer, may complicate the interpretation of the SMR. Pt is near the Stoner ferromagnetic instability and could become magnetic when in contact with ferromagnetic materials, as experimentally shown by x-ray magnetic circular dichroism (XMCD), anomalous Hall resistance (AHR), spin pumping, and first principle calculations of the Pt/YIG hybrid~\cite{Huang2012, Lin2013, Sun2013, Lu2013, Qu2013}. In order to separate the MPE form the pure SMR, many attempts have been made. By inserting a layer of Au or Cu between NM and FMI, the MPE can be effectively screened, but the SMR amplitude is largely suppressed as well~\cite{Miao2014,Althammer2013}. Furthermore, the insertion of an extra layer would introduce an additional interface whose quality is not easily accessible. An alternative approach to pursue the pure SMR is to find proper NM metals in direct contact with YIG, but without the MPE. The Au has a SOC strength comparable to Pt or Pd and it is free of the MPE, but it has an extremely weak inverse spin-Hall voltage and SMR~\cite{Qu2013, Wang2014}. According to the theoretical calculation~\cite{Tanaka2008}, the $4d$ metal Rh also possesses a large SOC strength and spin-Hall conductivity, and a small magnetic susceptibility, implying an insignificant MPE in the Rh metal when in contact with ferromagnetic materials. Thus Rh might be an excellent material for the pure SMR study.

In this article, the anisotropic magnetoresistance (AMR) and AHR of Rh/YIG and Pt/YIG hybrids were investigated in the temperature range of 5-300 K. Indeed, we show that the differences in  magneto-transport properties between these two hybrids are attributed to the strong (Pt) and weak (Rh) MPE, and thus, Rh/YIG provides an ideal model system for pure spin-current investigations.

\section*{Results}

\begin{figure}[tbp]
     \begin{center}
     \includegraphics[width=4in,keepaspectratio]{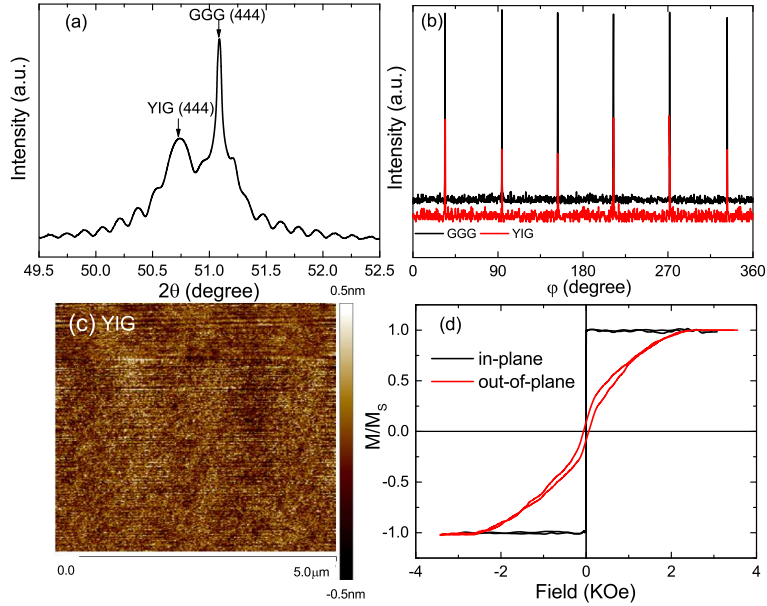}
     \end{center}
     \caption{(Color online) (a) A representative 2$\theta$-$\omega$ XRD patterns for YIG/GGG film near the (444) peaks of GGG substrate and YIG film. (b) The $\varphi$-scan of YIG/GGG film. (c) AFM surface topography of a representative YIG film. (d) The field dependence of normalized magnetization for YIG/GGG film measured at room temperature. For the in-plane (out-of-plane) magnetization, the magnetic field is applied parallel (perpendicular) to the film surface.}
     \label{fig1}
\end{figure}

Figure 1(a) plots a representative room-temperature 2$\theta$-$\omega$ XRD scan of epitaxial YIG/GGG thin film near the (444) reflections of gadolinium gallium garnet (GGG) substrate and YIG film. Clear Laue oscillations indicate the flatness and uniformity of the epitaxial YIG film. The epitaxial nature of YIG film was characterized by $\varphi$-scan measurements with a fixed 2$\theta$ value at the (642) reflections, as shown in Fig. 1(b). In this study, the thicknesses of YIG and Rh or Pt films, determined by fitting the x-ray reflectivity (XRR) spectra, are approximately 50 nm and 3 nm, respectively. The AFM surface topography of a representative YIG film in Fig. 1(c) reveals a surface roughness of 0.15 nm, indicating atomically flat of the epitaxial YIG film. As shown in Fig. 1(d), the in-plane and out-of-plane coercivities of the YIG film are $<$ 1 Oe and 60 Oe, respectively. The paramagnetic background of the GGG substrate has been subtracted and the magnetization is normalized to the saturation magnetization $M_\textup{s}$. The out-of-plane magnetization saturates at a field above 2.2 kOe, which is consistent with previous results~\cite{Lin2014,Huang2012}. The above properties indicate the excellent quality of our epitaxial YIG film.

\begin{figure}[tbp]
\begin{center}
\includegraphics[width=3.4in,keepaspectratio]{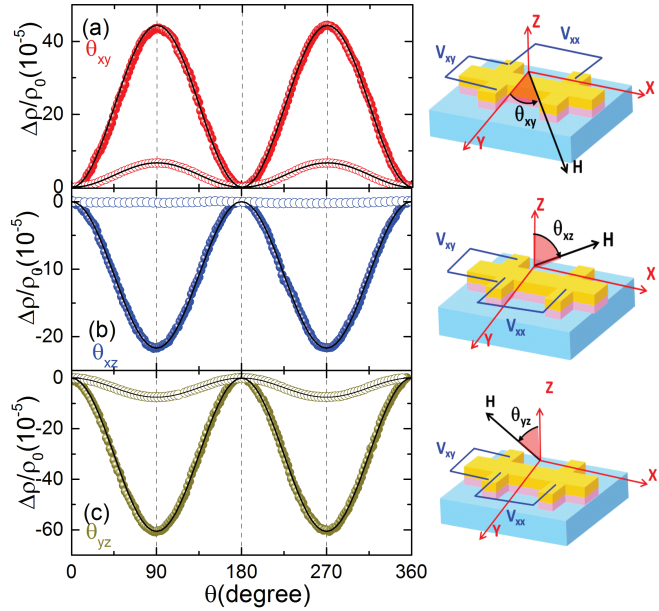}
\end{center}
\caption{Anisotropic magnetoresistance for the Rh/YIG (open symbols) and Pt/YIG (closed symbols) hybrids with the magnetic field scanning within the $xy$ (a), $xz$ (b), and $yz$ (c) planes. The AMR is measured at room temperature in a field of $\mu_0$H = 20 kOe. The solid lines through the data are fits to $\textup{cos}^{2} \theta$ with a 90 degree phase shift. The right panels show the schematic plots of longitudinal resistance and transverse Hall resistance measurements and notations of different field scans in the patterned Hall bar hybrids. The $\theta_{xy}$, $\theta_{xz}$, and $\theta_{yz}$ denote the angles of the applied magnetic field relative to the $y$-, $z$-, and $z$-axes, respectively.}
\label{fig2}
\end{figure}

Figures 2(a)-(c) plot the room-temperature AMR for the Rh/YIG (open symbols) and Pt/YIG (closed symbols) hybrids. As shown in the right panels, the Rh/YIG and Pt/YIG hybrids are patterned into Hall-bar geometry and the electric current is applied along the $x$-axis. The AMR is measured in a magnetic field of 20 kOe, which is sufficiently strong to rotate the YIG magnetization in any direction. Here the total AMR is defined as $\Delta$$\rho$/$\rho_0$ = [$\rho$(M $\parallel$ I) - $\rho$(M $\perp$ I)]/$\rho_0$. We note that when the magnetic field scans within the $xy$ plane [Fig. 2(a)], both the CAMR and SMR contribute to the total AMR, and it is difficult to separate them from each other; for the $xz$ plane [Fig. 2(b)], the magnetization of YIG is always perpendicular to the spin polarization of the spin current and the SMR is absent, and the resistance changes are attributed to the MPE-induced CAMR. For the $yz$ plane [Fig. 2(c)], the electric current is always perpendicular to the magnetization, the CAMR is zero, and only the SMR survives. According to Eq. (1), the amplitudes of CAMR or SMR ($\Delta$$\rho$/$\rho_0$) oscillate as a function of $\textup{cos}^{2} \theta$, as shown by the solid black lines in Fig. 2. Both the Rh/YIG and Pt/YIG hybrids display clear SMR at room temperature, with the amplitudes reaching 7.6 $\times$ 10$^{-5}$ and 6.1 $\times$ 10$^{-4}$, respectively [see Fig. 2(c)]. On the other hand, the CAMR also exists in the Pt/YIG, and its amplitude of 2.2 $\times$ 10$^{-4}$ is comparable to the SMR. However, as shown in Fig. 2(b), for Rh/YIG hybrid, the $\theta_{xz}$ scan shows negligible AMR and the resistivity is almost independent of $\theta_{xz}$, indicating the extremely weak MPE at the Rh/YIG interface in contrast to the significant effect at the Pt/YIG interface. The MPE at Pt/YIG interface was previously evidenced from the measurements of XMCD, AHR, and spin pumping~\cite{Huang2012,Sun2013,Lu2013}.

\begin{figure}[tbp]
     \begin{center}
     \includegraphics[width=4in,keepaspectratio]{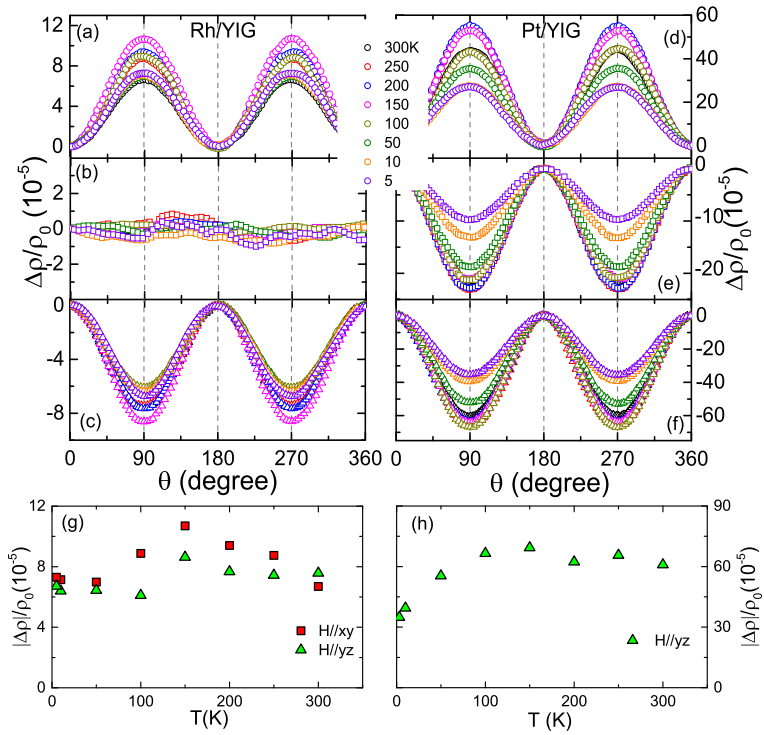}
     \end{center}
     \caption{Anisotropic magnetoresistance for the Rh/YIG hybrid at various temperatures down to 5 K for the $\theta_{xy}$ (a), $\theta_{xz}$ (b), and $\theta_{yz}$ (c) scans. The results of Pt/YIG are shown in (d) - (f). The AMR is measured in a field of $\mu_0$H = 20 kOe. (g) and (h) plot the temperature dependence of SMR amplitude for the Rh/YIG and Pt/YIG hybrids, respectively. The cubic and triangle symbols stand for the $\theta_{xy}$ and $\theta_{yz}$  scans, respectively.}
     \label{fig3}
\end{figure}

Upon decreasing temperature, the SMR persists down to 5 K in both the Rh/YIG and Pt/YIG hybrids [see Fig. 3]. However, there is no sizable CAMR in the Rh/YIG hybrid down to the lowest temperature [see Fig. 3(b)], indicating the extremely weak MPE at the interface even at low temperature. While for the Pt/YIG hybrid, as shown in Fig. 3(e), the amplitude of CAMR is almost independent of temperature  for $T > 100$ K, and then decreases by further lowering temperature, with the amplitude reaching 1.0 $\times$ 10$^{-4}$ at 5 K. The above features are quite different from the Pd/YIG hybrid, where the amplitude of CAMR increases as the temperature decreases, showing a comparable value to the SMR at 3 K~\cite{Lin2014}. The reason for these different behaviors is unclear, and further investigations are needed. Since the CAMR is negligible in Rh/YIG, the SMR dominates the AMR when the magnetic field is varied within the $xy$ plane. Figures 3(g) and (h) plot the temperature dependence of SMR amplitude for the Rh/YIG and Pt/YIG, respectively. The SMR amplitudes exhibit strong temperature dependence, reaching a maximum value of 1.1 $\times$ 10$^{-4}$ (Rh/YIG) and 6.9 $\times$ 10$^{-4}$ (Pt/YIG) around 150 K. Such nonmonotonic temperature dependence of SMR amplitude was previously reported in Pt/YIG hybrid, which can be described by a single spin-relaxation mechanism~\cite{Marmion2014}. It is noted that the hybrids with different Rh thicknesses exhibit similar temperature dependent characteristics with different numerical values compared to the Rh(3 nm)/YIG hybrid shown here. For example, the Rh(5 nm)/YIG hybrid reaches its maximum SMR amplitude of 0.8 $\times$ 10$^{-4}$  around 100 K.

In order to characterize the MPE at the NM/FMI interface, we also carried out the measurements of transverse Hall resistance R$_{xy}$ with a perpendicular magnetic field up to 70 kOe, as shown in Figs. 4(a) and (b). In both Rh and Pt thin films, the ordinary-Hall resistance (OHR), which is proportional to the external field, is subtracted from the measured R$_\textup{xy}$, i.e., R$_\textup{AHR}$ = R$_{xy}$ - R$_\textup{OHR}$$\times \mu_0H$, R$_\textup{AHR}$ is the anomalous Hall resistance. The resulting R$_\textup{AHR}$ as a function of magnetic field for the Rh/YIG and Pt/YIG hybrids are shown in Figs. 4(c) and (d), respectively. The AHR is proportional to the out-of-plane magnetization, and thus provides a notion for MPE at the NM/FMI interface. At room temperature, for the Rh/YIG hybrid, the R$_\textup{AHR}$ = 0.57 m$\Omega$, which is 22 times smaller than the Pt/YIG hybrid, implying the extremely weak MPE at Rh/YIG interface, being consistent with the AMR results in Fig. 2(b). We note that the R$_\textup{AHR}$ of Rh/YIG hybrids with different Rh thicknesses was also measured. For example, the R$_\textup{AHR}$ reaches 1.65 m$\Omega$ and 0.26 m$\Omega$ in Rh(2 nm)/YIG and Rh(5 nm)/YIG at room temperature, respectively. The temperature dependence of R$_\textup{AHR}$ for the Rh/YIG and Pt/YIG hybrids are summarized in Figs. 4 (e) and (f), respectively. As can be seen, the R$_\textup{AHR}$ exhibits significantly different behaviors: the R$_\textup{AHR}$ roughly decreases on lowering temperature in Rh/YIG. However, in Pt/YIG, the magnitude of R$_\textup{AHR}$ decrease with temperature for $T >$ 50 K and then it suddenly increases upon further decreasing temperature. Moreover, the R$_\textup{AHR}$ of Pt/YIG changes sign below 50 K, while it is stays positive for Rh/YIG. Similar non-trivial AHR were also observed in Pt/LCO hybrids~\cite{Tian2015}, but there is no existing quantitative theory to compare these results, further theoretical and experimental investigations are needed to clarify the dominating mechanisms.

\begin{figure}[tbp]
     \begin{center}
     \includegraphics[width=4in,keepaspectratio]{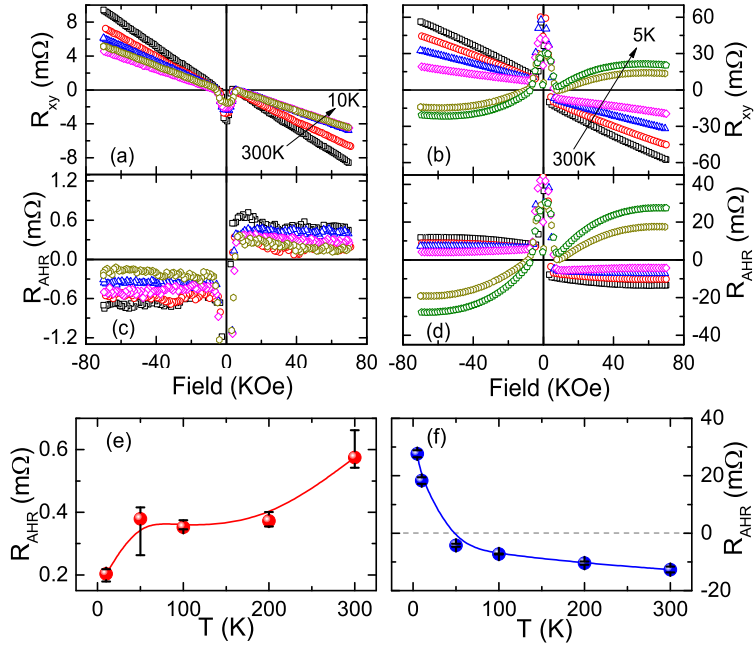}
     \end{center}
     \caption{Transverse Hall resistance R$_{xy}$ for the Rh/YIG (a) and Pt/YIG (b) hybrids as a function of magnetic field up to 70 kOe at different temperatures. The anomalous Hall resistance R$_\textup{AHR}$ for the Rh/YIG (c) and Pt/YIG (d) at different temperatures. The R$_\textup{AHR}$ can be derived by subtracting the linear background of OHR. Temperature dependence of R$_\textup{AHR}$ for the Rh/YIG (e) and Pt/YIG (f). All R$_\textup{AHR}$ are averaged by [R$_\textup{AHR}$(70 kOe)- R$_\textup{AHR}$(-70 kOe)]/2. The error bars are the results of subtracting OHR in different field ranges.}
     \label{fig4}
\end{figure}

\section*{Summary}

In summary, we carried out measurements of angular dependence of magnetoresistance and transverse Hall resistance in Rh/YIG and Pt/YIG hybrids. Both hybrids exhibit SMR down to very low temperature. The observed AHR and CAMR indicate a significant MPE at the Pt/YIG interface, while it is negligible at the Rh/YIG interface. Our findings suggest that the absence of the MPE makes the Rh/YIG bilayer system an ideal playground for pure spin-current related phenomena.

\section*{Methods}

The Rh/YIG and Pt/YIG hybrids were prepared in a combined ultra-high vacuum (10$^{-9}$ Torr) pulse laser deposition (PLD) and magnetron sputter system. The high quality epitaxial YIG thin films were grown on (111)-orientated single crystalline GGG substrate via PLD technique at 750 $^\circ$C. The thin Rh and Pt films were deposited by magnetron sputtering at room temperature. All the thin films were patterned into Hall-bar geometry. The thickness and crystal structure were characterized by using Bruker D8 Discover high-resolution x-ray diffractometer (HRXRD). The thickness was estimated by using the software package LEPTOS (Bruker AXS). The surface topography and magnetic properties of the films were measured in Bruker Icon atomic force microscope (AFM) and Lakeshore vibrating sample magnetometer (VSM) at room temperature. The measurements of transverse Hall resistance and longitudinal resistance were carried out in a Quantum Design physical properties measurement system (PPMS-9 T) with a rotation option in a temperature range of 5-300 K.

\section*{Acknowledgments}
We acknowledge the fruitful discussions with  S. M. Zhou. This work is financially supported by the National Natural Science Foundation of China (Grants No. 11274321, No. 11404349, No. 11174302, No. 51502314, No. 51522105) and the Key Research Program of the Chinese Academy of Sciences (Grant No. KJZD-EW-M05). S. Zhang was partially supported by the U. S. National Science Foundation (Grant No. ECCS-1404542).

\section*{Author contributions}

Q. F. Z., S. Z., and R. W. L. planned the experiments. T. S., L. M., and Y. L. X. synthesized the hybrids. Structure characterization, magnetic and transport measurements were performed by T. S., H. L. Y., Z. H. Z., H. H. L., and L. P. L.
The data were analysed by T. S., H. L. Y., Y. H. W., B. M. W., Q. F. Z., S. Z., and R. W. L. T. S., Q. F. Z., and S. Z. wrote the paper. All authors participated in discussions and approved the submitted manuscript.

\section*{Additional information}
\textbf{Competing financial interests:} The authors declare no competing financial interests.

\end{document}